\documentclass{iucrjournals}
\nolinenumbers
\usepackage{color,soul}
\usepackage{subcaption}
\usepackage{amssymb}
\title{Local order in natural sanidine disentangled via single crystal diffuse scattering}

\author[a,b]{Christin Wiggers \IUCrEmaillink{cwiggers@marum.de}}%
\author[c]{Daniel A. Chaney \IUCrEmaillink{daniel.chaney@esrf.fr}\IUCrOrcidlink{0000-0001-8582-3070}}%
\author[a,b,d]{Ella M. Schmidt \IUCrCemaillink{ella.schmidt@uni-bremen.de}\IUCrOrcidlink{0000-0002-0687-9963}}%

\affil[a]{MARUM - Center for Marine Environmental Sciences, University of Bremen, 28359 Bremen, Germany}
\affil[b]{Faculty of Geosciences, University of Bremen, Klagenfurter Stra\ss e 2-4, DE-28359 Bremen, Germany}
\affil[c]{ESRF, The European Synchrotron, 71 avenue des Martyrs, 
Grenoble, F-38043, France}
\affil[d]{MAPEX Center for Materials and Processes, University of Bremen, 28359 Bremen, Germany}

\begin{document} 
	\maketitle 
	
\begin{synopsis}
	A combined diffuse-scattering, 3D-$\Delta$PDF and atomistic-modelling study reveals that bowtie-like diffuse scattering in natural sanidine arises from coupled K/Na occupational disorder and anisotropic framework relaxations.
\end{synopsis}
	
\begin{abstract}
	Natural sanidine commonly preserves metastable chemical disorder at ambient conditions, but the local structural correlations associated with this disorder remain poorly understood due to a reliance on average-structure analysis alone. Here, we investigate single-crystal diffuse scattering in a natural sanidine
	from Drachenfels, Germany, with consistent observations across additional
	samples from Laacher See, Germany, and Vesuvio, Italy. The diffraction data reveal pronounced anisotropic diffuse features, including characteristic bowtie-like intensity distributions in reciprocal space. To resolve the origin of these features, we combine three-dimensional difference pair distribution function (3D-$\Delta$PDF) analysis with force-field calculations and Monte Carlo modelling. The 3D-$\Delta$PDF shows correlated deviations from the average structure and indicates a strong coupling between occupational and displacive disorder. Force-field calculations demonstrate that the aluminosilicate framework responds strongly to the local alkali occupancy, with the largest structural variation occurring along the $a$ direction. Guided by these observations, we construct a disorder model in which K/Na occupational short-range order is combined with chemically constrained Al/Si distributions and local framework relaxations. The model reproduces the main experimental diffuse-scattering features and shows that the bowtie-like diffuse scattering arises from correlated framework distortions driven primarily by local K/Na order.
\end{abstract}
	

	\section{Introduction}
	
	Feldspars are among the most abundant rock-forming minerals in the Earth's crust and occur in igneous \cite{brown1994feldspars}, metamorphic \cite{collerson1976composition}, and sedimentary \cite{morad1978feldspars} lithologies as well as in extraterestial materials, such as Martina meteorites \cite{rubin2017meteoritic}. Their structural evolution as a function of composition, temperature and pressure has therefore been investigated extensively \cite{kroll1991si,hovis1999high,pakhomova2020polymorphism}. Within the alkali feldspar series (NaAlSi$_3$O$_8$--KAlSi$_3$O$_8$), sanidine is the high-temperature monoclinic ($C2/m$) polymorph. At high temperatures the potassium endmember of the K-feldspars (KAlSi$_3$O$_8$) forms a solid solution with the sodium (NaAlSi$_3$O$_8$) endmember and sanidine often retains significant amounts of sodium due to rapid cooling. 
		
\begin{figure}[ht] %
		\begin{center}
			\includegraphics[height=0.4\textwidth]{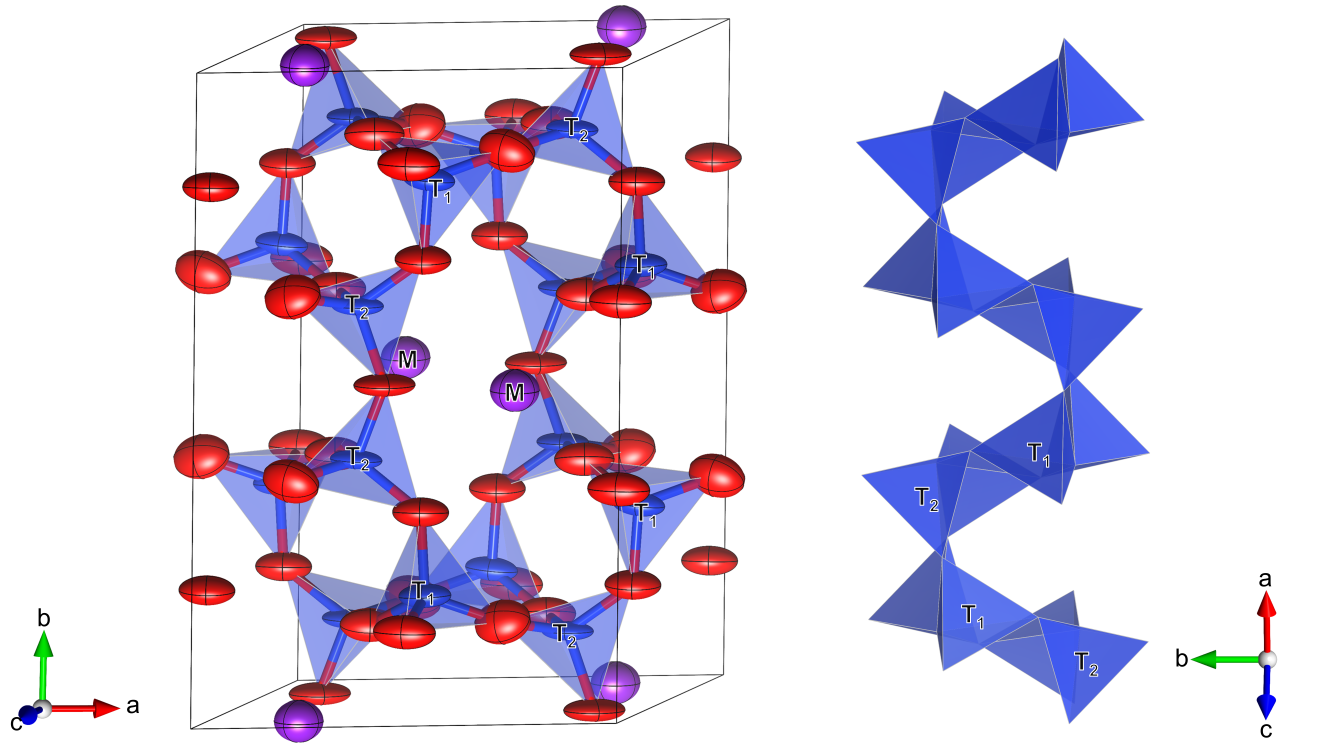} 
		\end{center}
		\caption{Right: Average structure model of K-feldspar. K in purple, O in red. Al/Si in blue. Atoms are shown with their anisotropic displacement parameters at 90~\% probability. $T_1$ and $T_2$ mark the two distinct tetrahedral sites and M the site occupied by the cation. Right: Double crankshaft chains of tetrahedra. } 
		\label{fig:figure1}
	\end{figure}

	The sanidine structure contains two crystallographically nonequivalent tetrahedral sites, $T_1$ and $T_2$, occupied by Al and Si, plus a larger irregular $M$ site occupied by alkali cations. The corner-sharing tetrahedra form four-membered rings and double crankshaft chains parallel to the $a$ axis (Fig.~\ref{fig:figure1}). In the classical description of the alkali feldspar order--disorder series, sanidine represents the relatively disordered high-temperature state, orthoclase shows partial Al/Si ordering with a preference for Al at $T_1$, and microcline is the low-temperature triclinic polymorph with strongly developed tetrahedral order. The alkali cations may also exhibit positional or occupational disorder within the framework cavities, as suggested by earlier crystallographic studies \cite{megaw_order_1959,ribbe_crystal_1994}. As the process of Al/Si reordering in feldspars onsets at temperatures above 500 °C \cite{brown_alkali_1989}, it is possible to investigate the metastable sanidine phases at ambient conditions.
	
		The sanidine solid solution series, therefore exhibits two chemical degrees of freedom in its average crystal structure: the explicit decoration of (a) Al and Si on the $T$- framework sites and (b) Na and K over the $M$ sites. Especially for (a) it is assumed that L\"owenstein's rule should be fulfilled: i.e.\ any explicit decoration of Al and Si over the $T$ sites should avoid Al-O-Al configurations due to the under-bonding of the corner-sharing oxygen \cite{loewenstein1954distribution}. However, the local occupational arrangement of these species and its coupling to framework relaxation cannot be resolved from Bragg diffraction alone. Single-crystal diffuse scattering is well suited to probing such local correlations \cite{welberry2016one,withers2005disorder,osborn2025diffuse,keen2015crystallography,fischer2025accessing}. Previous work has attributed diffuse scattering in sanidine mainly to distortions of the aluminosilicate framework associated with Al/Si disorder and the strain fields resulting from the different ionic radii of Si$^{4+}$ and Al$^{3+}$  \cite{pleger_diffuse_1996}. The extent to which local Na/K ordering contributes to this diffuse scattering, and how it couples to framework displacements, remains unclear.

		Here, we investigate local order in a natural sanidine sample from Drachenfels, Germany; similar diffuse-scattering patterns were also observed for sanidine samples from Laacher See, Germany, and Vesuvio, Italy. Using high-quality single-crystal diffuse scattering, three-dimensional difference pair distribution function (3D-$\Delta$PDF) analysis, force-field calculations and Monte Carlo modelling, we show that local K/Na occupational correlations drive anisotropic framework distortions that give rise to characteristic bowtie-like diffuse scattering. Our results clearly unravel the interplay between occupational and displacive disorder in sanidine and provide a structural model for local order in this feldspar.

	\section{Experimental and Computational Methods}
	
\subsection{Single-crystal diffuse scattering}
	
	High-quality single-crystal diffuse-scattering data were collected at the dedicated diffuse-scattering instrument on beamline ID28 at the European Synchrotron Radiation Facility (ESRF), Grenoble, France. Single crystals were polished to approximate cuboid shape with dimensions of about $200 \times 200 \times 400~\mu$m$^3$ and a surface roughness below $1~\mu$m. Each crystal was mounted on a $200~\mu$m MiTeGen loop using non-drying immersion oil (Type NVH, Cargille Laboratories).
	
	Measurements were carried out at ambient conditions using monochromatic X-ray radiation with $\lambda = 0.6968$~\AA. A full $360^\circ$ $\phi$ scan was recorded at detector angles of $19^\circ$ and $48^\circ$ using a PILATUS3 X 1M detector operated in shutterless mode. The sample was rotated continuously during acquisition, and frames were integrated over angular intervals of $0.25^\circ$ with an exposure time of 1~s. Corresponding measurement details for the additional sanidine samples are provided in the Supporting Information.
	
	Bragg reflections were indexed using CrysalisPro \cite{crysalis} to confirm phase identification based on the average structure reported by \citeasnoun{ohashi1974refinement}. For average-structure determination, the same scan geometry was used with shorter exposure times (0.1 s), finer angular integration steps ($0.1^\circ$), and an attenuation factor of $\approx$ 15 to prevent overexposure of strong Bragg reflections.
	
\subsection{Average structure refinement}
The average structure was solved with SHELXT and refined with SHELXL~\cite{sheldrick2008short} as implemented in \textit{Olex2}~\cite{dolomanov2009olex2}. Inspection of the refined $T$--O bond lengths indicated the expected average tetrahedral occupancies. The Al/Si occupancies of the $T$ sites were therefore fixed at 0.25 and 0.75, respectively, since the X-ray scattering contrast between Al and Si is limited and no robust site-specific deviation from the nominal average occupancy was indicated by the bond lengths.  In contrast, the occupational fraction $x$ of the $M$ site was refined freely, constraining the overall occupancy to K$_{x}$Na$_{1-x}$.

The refinement converged normally and the resulting average-structure model is provided in the Supporting Information, where further details of the data collection and refinement are also given.

\subsection{3D-$\Delta$PDF}

The 3D-$\Delta$PDF was obtained by Fourier transformation of the diffuse scattering after removal of the Bragg intensities. Bragg peaks were identified using a local median filter combined with a median absolute deviation (MAD) criterion and were replaced by values derived from the local median and MAD. \cite{weng2020k}. To suppress termination effects and reduce Fourier ripples, the diffuse intensities were multiplied by a three-dimensional Gaussian damping function before transformation.

The resulting 3D-$\Delta$PDF maps show deviations from the average structure in real space. Negative features correspond to interatomic vectors that occur less frequently than in the average structure, while positive features indicate vectors that occur more frequently.
 		
 \subsection{Structures relaxed with force-field models}
 \label{sec:FFMethods}
 To investigate the coupling between occupational disorder and structural relaxation, we performed force-field calculations using the General Utility Lattice Program (GULP) \cite{gale1997gulp,gale2003general}. Starting from the K endmember structure reported by \citeasnoun{tseng1995characterization} (ICSD code 81137), all distinct occupational variants of a $1 \times 1 \times 1$ unit cell with composition KAlSi$_3$O$_8$ were generated using the \textit{supercell} program \cite{okhotnikov2016supercell}. For the tetrahedral framework, this corresponds to distributing 2 Al and 6 Si atoms over the available $T_1$ and $T_2$ sites. Among the 111 possible Al/Si arrangements, only those satisfying Löwenstein's rule were retained, yielding 8 symmetry-independent framework configurations without Al--O--Al linkages.
 
 For the alkali site, 16 distinct K/Na occupational configurations were considered within the $1 \times 1 \times 1$ cell. These include both the full compositional range from 4 K + 0 Na to 0 K + 4 Na and the different explicit arrangements of K and Na over the nominally symmetry-equivalent $M$ sites. Combining the 8 framework configurations with the 16 alkali configurations gave a total of 128 unique unit-cell models.
 
 All models were relaxed in space group $P1$ to account for the symmetry reduction introduced by the explicit occupational configurations. Lattice energies were minimized using a combination of Buckingham, three-body, and core--shell spring potentials \cite{dove1997lattice,binks1994computational}. The potential parameters are given in the Supporting Information.
 		
 	\subsection{Supercells for modelling diffuse scattering}
 	\label{sec:MethodModel}
 
 	The experimentally observed diffuse scattering was modelled using $20 \times 20 \times 20$ supercells with a Si:Al ratio of 75:25 on the tetrahedral ($T$) sites and a K:Na ratio of 50:50 on the metal ($M$) site. The final configurations were generated in a multistep procedure, guided by the experimental diffuse scattering, 3D-$\Delta$PDF analysis, force-field calculations (\ref{sec:FFMethods}), and basic crystal chemical constraints. Full discussion of the reasoning underpinning our modeling approach can be found in section~\ref{sec:DisorderModel}.
 	
 	\subsubsection{Step 1: Al/Si order}
 	A starting configuration was employed where all $T$/$M$ sites are fully occupied by Si/K, respectively, from which 25~\% of Si were randomly replaced by Al. Overall composition of exactly Si:Al = 75:25 was maintained to satisfy the charge-balanced framework stoichiometry. Al/Si occupational disorder was introduced by Monte Carlo simulation using DISCUS \cite{neder2008diffuse}, where chemically distinct atoms occupying the same crystallographic site are allowed to swap and all pairs of $T$-sites connected through a common corner-sharing oxygen atom were treated as nearest neighbours. The swaps were guided by a target Warren--Cowley short-range-order (SRO) parameter $\alpha_{\mathrm{Al}}$ \cite{warren1951atomic}, defined as
 	\begin{equation}
 		\alpha_{\mathrm{Al}} = 1 - \frac{p_{\mathrm{Al,Si}}}{m_{\mathrm{Al}} m_{\mathrm{Si}}},
 	\end{equation}
 	where $p_{\mathrm{Al,Si}}$ is the conditional probability of finding an Al atom on a $T$ site adjacent to a Si atom on a neighbouring $T$ site, and $m_{\mathrm{Al}}$ and $m_{\mathrm{Si}}$ are the molar fractions of Al and Si, respectively. To enforce Löwenstein's rule, a target value of $\alpha_{\mathrm{Al}} = -1$ was imposed, thereby maximizing Al--O--Si and excluding Al--O--Al linkages.
 	
 	\subsubsection{Step 2: K/Na order}
 	
 	The K/Na occupational arrangement was modelled using the disordered superspace approach \cite{schmidt2019interpretation} whereby local occupational correlations are described by two cosine-based modulation functions defining the probability of finding K at a given atomic position. In contrast to conventional superspace descriptions of long-range-ordered modulated structures, the use of two modulation functions introduces local order which disrupts strict periodicity. In DISCUS, the structural model initially contains placeholder atoms on all alkali positions to which the following probabilities were assigned 
 	
 	\begin{equation}
 		p^{\mathrm{K}}_{\pm} = 0.5 \pm 0.25 \cos \left[ 2\pi \left( {q}_x \cdot x + q_z \cdot z \right) \right].
 	\end{equation}
 	
 	Half of the placeholders being assigned to the $p^{\mathrm{K}}_{+}$ modulation and half to the $p^{\mathrm{K}}_{-}$ modulation with $\vec{q}$ determined from diffuse scattering data. A Monte Carlo simulation was then used to sort the placeholders into clusters considering the six nearest K-K interatomic vectors together with the [010] vector. The majority of which used a target Warren-Cowley parameter of +1, favouring strong positive occupational correlations however, for the [010] and [001] vectors a reduced target value of 0.5 was employed to match the experimentally observed broadening of diffuse scattering features, as the placeholder cluster size determines the width of the diffuse maximum \cite{schmidt2019interpretation}. The simulation was run for 100 × N$_{atom}$ cycles at an effective temperature of 2.
 	
 	The relevant modulation function was then evaluated for each placeholder position. The resulting probability was compared with uniformly distributed random number between 0 and 1; if the probability exceeded the random number, the placeholder was replaced by K, otherwise by Na. This procedure yields a supercell with an overall K:Na ratio of 50:50 and the desired local occupational correlations.
 	
 	\subsubsection{Step 3: Relaxation of the supercell}
 	\label{sec:SupercellRelax}
 	Finally, the substitutionally disordered supercells were relaxed via a two step process using custom Fortran scripts to reproduce the framework distortions inferred from the force-field calculations. \\
 	Firstly strain was applied to each local unit cell based on its Ka/Na composition and the corresponding lattice parameters determined from the force-field model (see 2.4). This initial relaxation step was performed using a course-grained Monte Carlo approach detailed further in supplementary material. \\
 	In a second step, all AlO$_4$ and SiO$_4$ tetrahedra were replaced by tetrahedra with idealized bond lengths and bond angles, producing split oxygen positions. A subsequent Monte Carlo simulation allowed the tetrahedra to translate and rotate about their central $T$ atoms, while the K and Na atoms were also allowed to shift. The optimization employed the same Buckingham potentials as used in the GULP calculations to preserve realistic K--O and Na--O bond geometries. A core–shell spring potential was used to drive the split oxygen positions associated with neighbouring tetrahedra towards each other. This Monte Carlo simulation was run in a fortran program, for 500 × N$_{M,T}$ cycles at an effective temperature of 0, where N$_{M,T}$ is the total number of $M$- and $T$-site atoms in the supercell. After optimization, the split oxygen positions were merged by averaging their positional coordinates. 
 	
 	\subsubsection{Step 4: Diffuse scattering calculation}
 	To evaluate the diffuse scattering from the resulting disorder model, twenty $20 \times 20 \times 20$ supercells were generated via steps 1-3. Diffuse scattering was calculated for each configuration using DISCUS \cite{neder2008diffuse} on a reciprocal-space grid defined by $-10 \leq h,k,l \leq 10$ with $\Delta h = \Delta k = \Delta l = 0.05$. The calculated diffuse intensities were averaged over all configurations and symmetry-averaged for $2/m$ Laue symmetry using \textit{Meerkat} \cite{simonov2020meerkat}. The resulting dataset was then resampled using a Lanczos-type interpolation ($m = 4$) \cite{lanczos1964evaluation,paddison2019ultrafast}.
	
	\section{Results and Discussion}
	
	\subsection{Average structure}
	
	Refinement of the average structure yielded mixed alkali occupancy of approximately $\mathrm{K_{0.5}Na_{0.5}}$ for the sanidine crystal discussed in this study. A striking feature of the refinement is the behaviour of the atomic displacement parameters (ADPs). Whereas the alkali $M$ site exhibits comparatively small and nearly isotropic ADPs, the Al/Si--O framework atoms show markedly enlarged and anisotropic displacement ellipsoids. The principal anisotropy lies predominantly within the $ab$ plane, with the longest ellipsoid axes oriented approximately along $a$ (Fig.~\ref{fig:figure1}).
	
	This pronounced anisotropy indicates substantial unresolved local displacements of the framework in the average structure. Such behaviour is consistent with local distortions required to accommodate chemical substitution on both the $M$- and $T$-sites. This ADP pattern already suggests we may expect to observe strongly directionally dependent diffuse scattering and suggests the presence of a framework-dominated displacive disorder coupled to compositional heterogeneity.
	
	\subsection{Diffuse scattering}
	\label{sec:DiffuseData}
	
\begin{figure}[ht] %
		\begin{center}
			\includegraphics[width=1.0\textwidth]{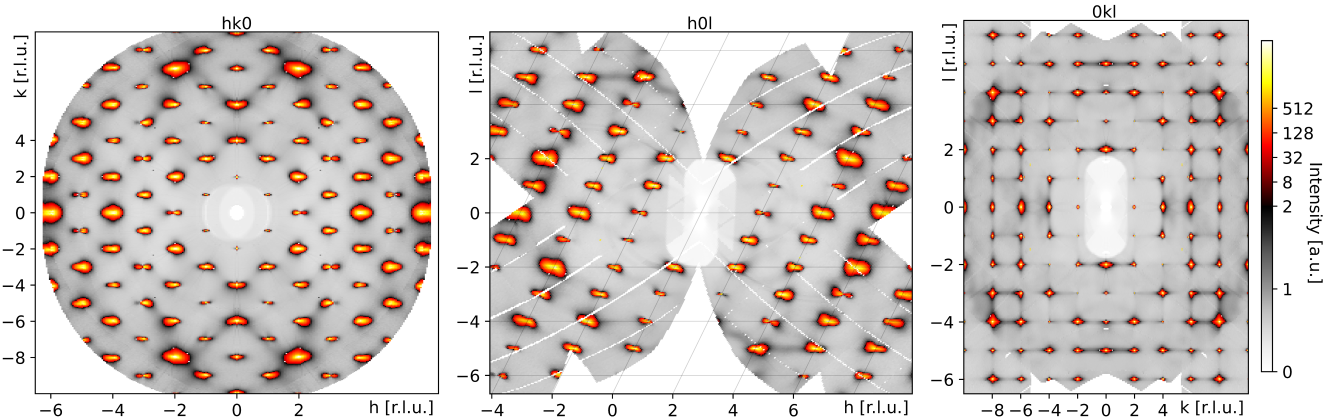} 
			\includegraphics[width=1.0\textwidth]{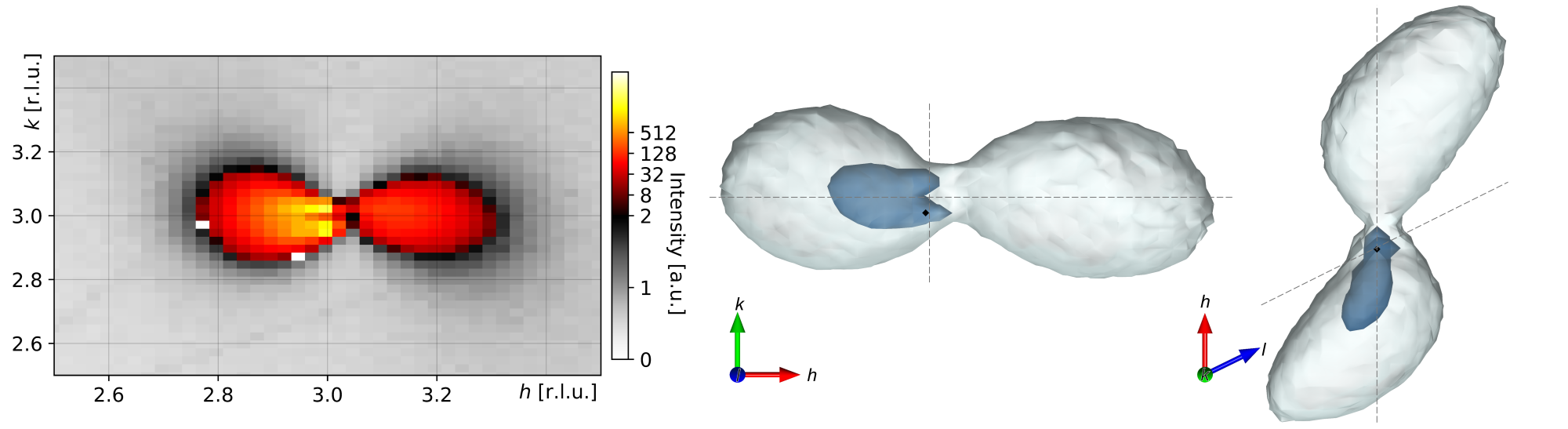}
		\end{center}
		\caption{Top: Symmetry-averaged diffuse-scattering intensity in the $(hk0)$, $(h0l)$, and $(0kl)$ reciprocal-space planes. A coordinate grid is superimposed on the $(h0l)$ section to aid visualization. Bottom: Close-up view of the bowtie-like diffuse-scattering feature surrounding the $(300)$ reflection (left), together with two three-dimensional renderings of the same reconstructed diffuse scattering viewed from different perspectives (middle and right). The dotted lines indicate the respective lattice directions, with the intersection at the (300) Bragg reflection, highlighting the off-centering of the bowtie.
		 } 
		\label{fig:figure2}
	\end{figure}

	Reconstructed three-dimensional reciprocal space reveals pronounced anisotropic diffuse-scattering features that, to our knowledge, have not previously been reported for sanidine. As shown in Fig.~\ref{fig:figure2}, the $hk0$ and $h0l$ sections exhibit characteristic bowtie-like diffuse intensity distributions adjacent to Bragg reflections such as $(200)$ and $(330)$. In both planes, the diffuse intensity forms teardrop-shaped lobes elongated predominantly along $h$. These features are displaced from the Bragg positions, such that the center of the bowtie does not coincide with the Bragg peak, and exhibit a pronounced asymmetry in intensity between the two lobes, with one lobe consistently displaying significantly higher intensity than the other (Fig.~\ref{fig:figure2}).
	
	The $h0l$ sections display motifs similar to those observed in $hk0$, although the diffuse lobes also display a small $l$ component. By contrast, the $0kl$ plane shows no bowtie-like scattering and is dominated by weaker, more streak-like diffuse features. These resemble the diffuse scattering previously reported for orthoclase and attributed to Al/Si tetrahedral disorder \cite{pleger_diffuse_1996}. The full three-dimensional reconstruction shows that the diffuse component is highly anisotropic and strongly direction-dependent.
	
	A general feature of the data is that the diffuse intensity increases with distance from the centre of reciprocal space. This behaviour is characteristic of displacement-dominated disorder and indicates that correlated atomic displacements make a major contribution to the observed scattering. At the same time, the diffuse intensity is concentrated in broad maxima that recur at similar positions within successive reciprocal-lattice units rather than being centred directly beneath the Bragg reflections. This pattern is consistent with a weakly developed, short-range modulation, which can be approximated by a modulation wavevector of $\vec{q} \approx \left(\frac{1}{9},0,-\frac{1}{20}\right)$. Representative examples of the two-dimensional Gaussian fits used to determine the $\vec{q}$ vector are provided in the Supporting Information.

	\subsection{3D-$\Delta$PDF}
	
\begin{figure}[ht] %
		\begin{center}
			\includegraphics[width=1.0\textwidth]{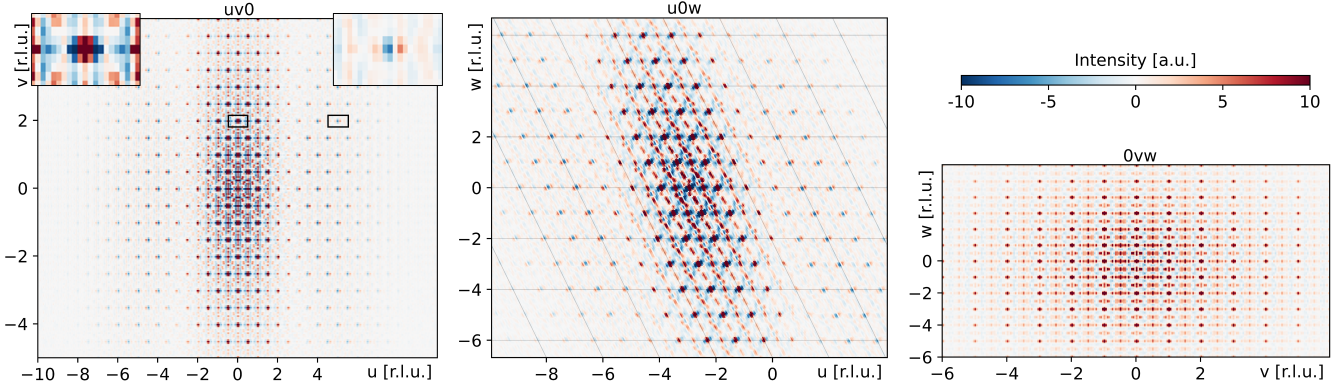} 
		\end{center}
		\caption{3D-$\Delta$PDF reconstructions in the $(uv0)$, $(u0w)$, and $(0vw)$ planes (left to right), generated from the diffuse scattering data in Figure~\ref{fig:figure2}. Negative and positive features, corresponding to interatomic vectors occurring less or more frequently than in the average structure, are coloured blue and red, respectively. Selected regions in the $(uv0)$ reconstruction are highlighted by rectangles and shown as enlarged views in the insets. A coordinate grid is included in the $(u0w)$ reconstruction as a guide to feature locations.} 
		\label{fig:figure3}
	\end{figure}

	The 3D-$\Delta$PDF, obtained by Fourier transformation of the diffuse scattering, provides a real-space view of the local correlations underlying the observed diffuse features~\cite{weber2012three}. The reconstructed maps contain maxima and minima at positions corresponding to interatomic vectors of the average structure, showing that the deviations from the average arrangement are highly correlated rather than random (Fig.~\ref{fig:figure3}).
	
	In the $(uv0)$ section, positive features dominate at short distances along $u$, approximately for $u \lesssim 2.5$, followed by a change in sign between about $u \approx 2$ and $3$, and by pronounced negative features centered near $u \approx 4$. Thus, correlations that are positive at short distances become negative at intermediate distances, and vice versa. A similar oscillatory behaviour is observed in the $(u0w)$ section, where the alternation of positive and negative features persists primarily along $u$. In contrast, the $(0vw)$ section is dominated mainly by positive correlations.
	
	This systematic alternation of positive and negative features over several unit cells is consistent with the interpretation of the diffuse scattering in terms of a weakly developed, short-range modulation with approximate wavevector $\vec{q} \approx \left(\frac{1}{9},0,-\frac{1}{20}\right)$. The modulation is not long-range periodic, but the 3D-$\Delta$PDF indicates that the local correlations nonetheless retain a clear directional characteristic.
	
	Closer inspection of selected profiles in the 3D-$\Delta$PDF reveals the characteristic ``Mexican hat'' line shape described by \citeasnoun{weber2012three}, which is an indicative signature of displacive disorder. In the present case, these features are interpreted as signatures of coupled occupational and displacive correlations. If atom pairs of the same type are displaced coherently in the same direction, a positive feature can appear at the average vector position, flanked by negative side lobes. Conversely, for opposing displacements, the central feature becomes negative and is surrounded by positive side lobes. The observed fine structure of the features in the 3D-$\Delta$PDF therefore indicate that occupational disorder is coupled to correlated atomic displacements (compare inset in Fig.~\ref{fig:figure3}).
	
	These signatures occur not only for vectors corresponding to unit-cell translations, but also at shorter interatomic vectors within the structure. For example, features near $\left(\frac{1}{2},0,0\right)$ include contributions from $T$--O vectors, whereas those near $\left(\frac{1}{2},\frac{1}{2},0\right)$ contain contributions from $M$--$M$ and $T$--$T$ correlations. The detailed shape of these features reflects the superposition of several pair contributions and illustrates the complexity of the coupled occupational and displacive disorder.

\subsection{Force-field-relaxed unit cells}

\begin{figure}[ht] %
	\begin{center}
		\includegraphics[height=0.3\textwidth]{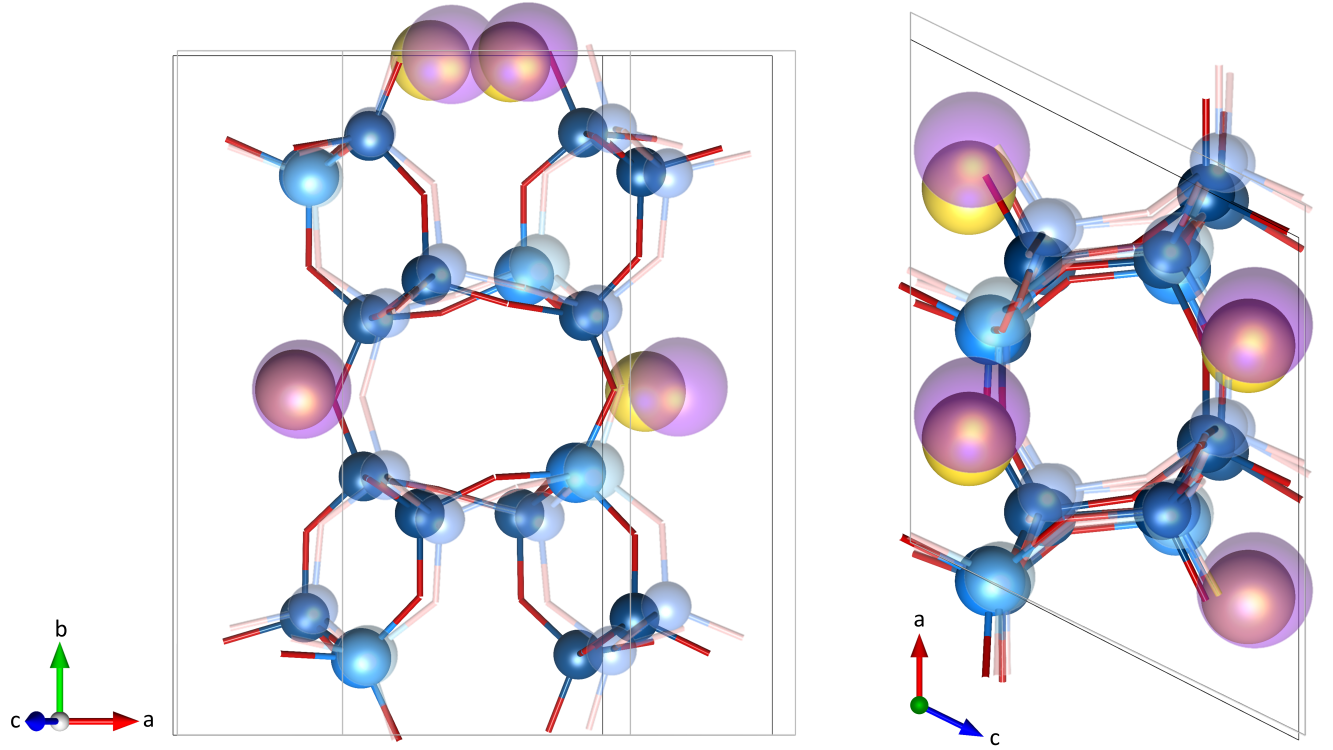} 
	\end{center}
	\caption{The unit cell of the relaxed K endmember is superimposed on the Na endmember, with purple and yellow spheres representing the respective cations. The framework consists of Al (light blue) and Si (dark blue) sites. For clarity, the K endmember is displayed with partial transparency, and both structures maintain the same Al/Si distribution.} 
	\label{fig:figure4}
\end{figure}

This complex coupling is further supported by our force-field calculations which show that the unit-cell geometry depends strongly on the alkali-site occupancy. Figure~\ref{fig:figure4} compares the relaxed structures of the Na and K endmembers and illustrates a pronounced elongation along $a$ in the K endmember; the $a$ lattice parameter is enlarged by $\approx$0.4~\AA. The $a$ lattice parameter also shows the largest variation when considering the full K--Na compositional range (see Supporting Information for the full list of lattice parameters or the relaxed structures). The largest incremental change in the simulated structures occurs upon substitution of the first K atom into the unit cell. The $b$ and $c$ lattice parameters also increase with increasing K content, but much less strongly, by only about 0.1~\AA{} and 0.06~\AA{}, respectively.
	
The cell angles vary less systematically. In particular, $\alpha$ spans a relatively broad range in Na-rich configurations, from approximately $87^\circ$ to $93^\circ$, reflecting the expected lower triclinic symmetry of the Na endmember. The observed geometric changes are governed primarily by the K/Na distribution; no systematic dependence on the explicit Al/Si arrangement was identified within the present set of force-field calculations.

The dominant structural response along $a$ is in good agreement with the anisotropic framework ADPs obtained from the average-structure refinement and with the directional diffuse-scattering features observed experimentally.

\subsection{Disorder model generation}
\label{sec:DisorderModel}
Taken together our diffuse-scattering observations, 3D-$\Delta$PDF analysis and force-field calculations define the essential ingredients of a disorder model that can both reproduce the experimental data and satisfy basic crystal-chemical constraints. In particular, the bowtie-like diffuse features suggest a weakly developed, short-range modulation, but the scattering is too broad and diffuse for a conventional superspace refinement of well-defined satellite reflections. We therefore constructed a real-space disorder model using the multistep Monte Carlo approach using the following rationale, for full details of computational implementation see Section~\ref{sec:MethodModel} and the Supporting Information.

\subsubsection{Al/Si disorder}

Given that the X-ray scattering contrast between Al and Si is weak, diffuse x-ray scattering does not allow a unique determination of Al/Si short-range order in this complex framework structure. We therefore impose the most robust crystal-chemical constraint, namely Löwenstein’s rule, and generate Al/Si configurations that avoid Al--O--Al linkages using a Monte Carlo approach. In this way, the tetrahedral disorder is treated in a chemically realistic manner without over-interpreting the limited direct experimental contrast.

\subsubsection{K/Na disorder}
The experimental diffuse scattering and the 3D-$\Delta$PDF are dominated by signatures of displacement disorder, but alkali-site occupancy is the most plausible origin of these distortions. This interpretation is strongly supported by the force-field calculations, which show a pronounced dependence of the unit-cell geometry on the K/Na distribution, especially along the $a$ direction, in agreement with the anisotropic ADPs from the average-structure refinement. Together with the alternating positive and negative features in the 3D-$\Delta$PDF at vectors involving the alkali sites, these observations indicate that K/Na occupational short-range order is the primary driver of the displacive disorder.

Because the diffuse maxima are strongly broadened relative to the Bragg reflections, we model the K/Na arrangement using the disordered superspace approach with an approximate modulation vector $\vec{q} \approx \left(\frac{1}{9},0,-\frac{1}{20}\right)$ to produce a supercell with overall 50:50 K/Na ratio with local occupational correlations that, on  a local level, respect the experimentally determined modulation vector. 
 
\subsubsection{Framework relaxation}
Our force-field calculations demonstrate that the tetrahedral framework responds strongly to the local K/Na arrangement on the $M$ sites. However, a simple local relaxation based only on tetrahedral rotations, tetrahedral translations and alkali displacements is not sufficient to reproduce the composition-dependent contraction and expansion of the unit cell. We therefore adopt the two step relaxation strategy. First, local unit cells are stretched or compressed according to their K/Na content based on trends derived from the force-field calculations. Second, the resulting structure is relaxed locally using Monte Carlo moves constrained by force-field-derived bond geometries. This procedure allows the model to capture both the local strain field imposed by alkali-site occupancy and the subsequent adjustment of the framework.
	
\subsection{Comparison of modeled and experimental diffuse scattering}

\begin{figure}[ht] %
	\begin{center}
		\includegraphics[width=1.0\textwidth]{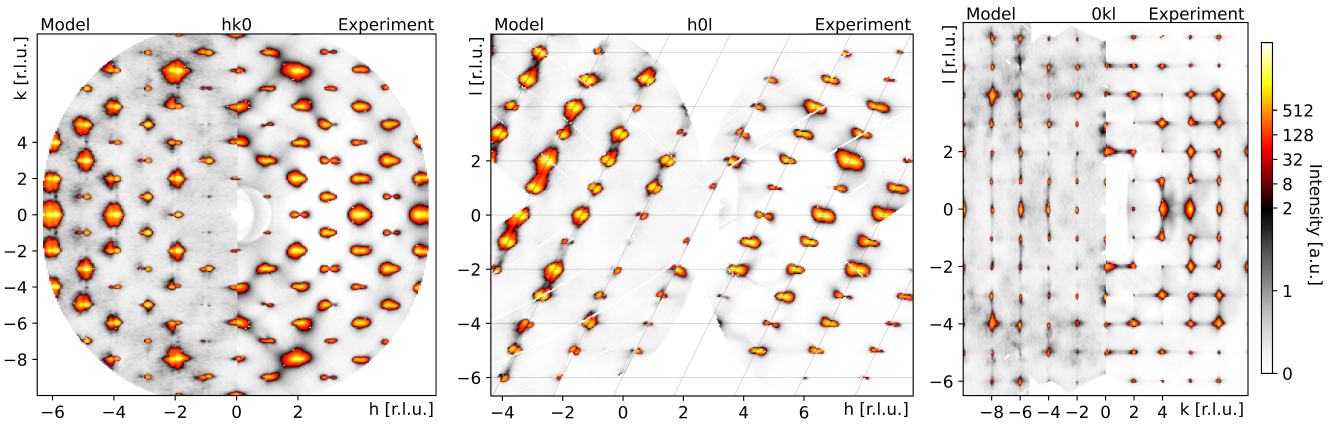} 
		\includegraphics[width=1.0\textwidth]{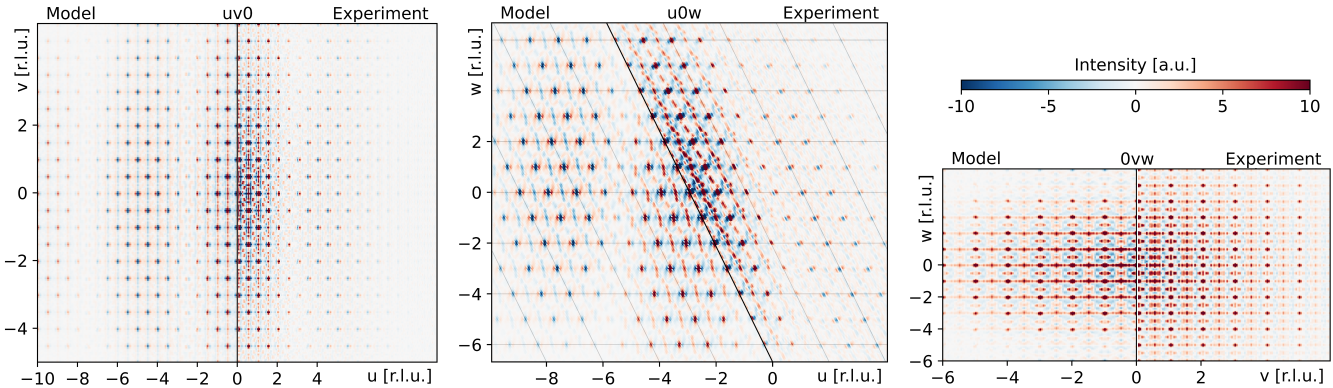}
	\end{center}
	\caption{Top: Comparison of experimental and modelled diffuse scattering in the $hk0$, $h0l$ and $0kl$ planes. The model reproduces the main anisotropic diffuse features, including the characteristic bowtie-like motifs in $hk0$ and $h0l$. Bottom: Comparison of experimental and modelled 3D-$\Delta$PDF in the $uv0$, $u0w$ and $0vw$ planes. The model reproduces the main  features, including the characteristic flip of intensities. A grid is superimposed on the $(h0l)$ and $(u0w)$ section to facilitate comparison of feature positions.}
	\label{fig:figure5}
	\end{figure}

The calculated diffuse scattering is shown in Figure~\ref{fig:figure5} and reproduces the principal features of the experimental data. In particular, the model captures the characteristic bowtie-like motifs in the $hk0$ plane, including the broad overlapping diffuse lobes around the Bragg reflections. The successful reproduction of this distinctive topology indicates that the model contains the essential ingredients of the underlying local disorder.

Some minor differences between model and experiment remain. In the reconstructed $hk0$ section, the bowtie-like diffuse features generally appear more pronounced in the experimental data. In some cases, diffuse intensity that is clearly visible in the experiment is only weakly developed in the model (compare (150) and ($\overline{1}50$)), whereas in others the relative intensity difference between opposing lobes is somewhat stronger in the simulation (compare (200) and ($\overline{2}00$)). Nevertheless, the model successfully reproduces the pronounced asymmetry observed for many of the bowtie features.
A more detailed comparison of line cuts along $h$ (see Supporting Information) reveals a similarly mixed picture. For some reflections, the simulated features are sharper and exhibit higher intensity than observed experimentally, whereas for others the experimental diffuse intensity exceeds that predicted by the model. There are also isolated cases in which the model predicts weak bowtie-like features that are not clearly resolved in the experimental data. At the same time, many reflections show very good agreement between model and experiment in both shape and intensity. Overall, these discrepancies are minor compared with the strong correspondence in the geometry, anisotropy and intensity distribution of the diffuse scattering.

The comparison in the $h0l$ plane further supports the validity of the model. Although the simulated features are slightly broader along $l$ and somewhat compressed along $h$, the relative intensity distribution agrees well with experiment. Specifically, the model precisely replicates the experimental intensity gradients between key reflections, when comparing ($200$) to ($\overline{2}00$) and ($400$) to ($\overline{4}00$). The agreement is also good in the $0kl$ plane, where the model reproduces the weak diffuse streaks along $l$. However, the simulation does not fully recover some of the weaker diffuse streaks observed experimentally along $k$. Overall, the comparison shows that the combination of K/Na occupational correlations and force-field-inspired framework relaxation is sufficient to explain the dominant diffuse-scattering features in natural sanidine.

The complementary comparison of the model and experimental 3D-$\Delta$PDFs yield an overall similar pictures. In all three layers shown in Fig.~\ref{fig:figure5} the dominating experimental features are correctly reproduced by the model. In the $uv0$ layer both the Mexican hat-like signal and the modulation of the 3D-$\Delta$PDF intensities along $u$ are reproduced. The model has a more pronounced retention of the modulation, with distinct maxima still visible at $u\approx-10$, while the experimental 3D-$\Delta$PDF features have decayed to a more significant degree, likely indicating that the modulation functions in the disordered superspace are ordered to a too large degree.  The $u0w$ layer also shows very good general agreement between model and data, with the only notable deviation being the more pronounced `tilt' of the Mexican hat signature in the experimental data compared to the model indicating that overall structural relaxations along $w$ may be slightly under-represented in the model. The $0vw$ layer shows similarly good agreement, but with modelled correlations that decay faster then those determined experimentally. 

\section{Conclusion and outlook}

Single-crystal diffuse scattering from natural sanidine reveals pronounced anisotropic diffuse features, most notably characteristic bowtie-like intensity distributions in reciprocal space. These features are not explained by the average structure alone, but require a description of correlated local disorder. By combining diffuse-scattering analysis, 3D-$\Delta$PDF, force-field calculations and Monte Carlo modelling, we show that the dominant local correlations arise from K/Na occupational short-range order coupled to anisotropic framework relaxations.

The force-field calculations demonstrate that the aluminosilicate framework responds strongly to the local alkali distribution, with the largest structural variation occurring along $a$. This behaviour is consistent with both the anisotropic atomic displacement parameters of the average structure and the directional character of the diffuse scattering. A disorder model combining chemically constrained Al/Si distributions, correlated K/Na arrangements and local framework relaxation reproduces the main experimental diffuse-scattering features and identifies local alkali order as the primary driver of the observed displacive disorder.

More generally, this study demonstrates that the interplay between occupational and displacive disorder in feldspars can be resolved by combining single-crystal diffuse scattering with real-space and atomistic modelling. The approach should be transferable to other sanidine samples and more broadly to compositionally complex framework silicates. Future work may address how the local-ordering behaviour evolves as a function of composition, temperature and pressure.
		
	\begin{acknowledgements}
	The authors acknowledge the help of Benjamin Fahl (ETH Z\"urich), Erik Neumann, Paul Benjamin Klar, Carla Uribe (all three University of Bremen), Giovanni O. Lepore (Florence University) and Carsten Paulmann (Uni Hamburg) during synchrotron measurements. We thank Alexei Bosak (ESRF) for fruitful discussion on the bowtie shape of the diffuse scattering. Christoph Vogt and Wolfgang Bach (University of Bremen) are acknowledged for fruitful discussions. C.W. acknowledges support by the MARUM Research Academy, University of Bremen.
	
	\end{acknowledgements}
	
	\begin{funding}
Funding was provided by the Bremen Cluster of Excellence ``The Ocean Floor—Earth’sUncharted Interface”. C.W. acknowledges support by GLOMAR – Bremen International Graduate School for Marine Sciences, University of Bremen.
The authors acknowledge the European Synchrotron Radiation Facility (ESRF) for provision of synchrotron radiation facilities under proposals ES-1706 and  ES-1502 on beamline ID28. We acknowledge DESY (Hamburg, Germany), a member of the Helmholtz Association HGF, for the provision of experimental facilities. Parts of this research were carried out at PETRA III and we would like to thank Carsten Paulmann for assistance in using P24. Beamtime was allocated for proposal  I-20231115
	\end{funding}
	
	\ConflictsOfInterest{There are no conflicts of interest to declare.
	}
	
	\DataAvailability{Raw data and a 3D reciprocal space reconstruction are made availabe at doi:0.5281/zenodo.21923941. Scripts for modelling can be obtained from the corresponding author upon reasonable request.
	}

	\bibliography{PaperSanidine} 

\end{document}